Preprinted manuscript

# A putative model of the gut-muscle axis in aged livestock

Karin Suzuki[2], Aoi Fukushima[2], Yu Adachi[2], Arisa Sano[1], Daisuke Yamamoto[1], Tsubasa Irie[1], Hirokuni Miyamoto[3,4,5,6,7], Shigeharu Moriya[8], Makiko Matsuura[4,6], Naoko Tsuji[6], Takashi Satoh[9], Tamotsu Kato[3,4], Takumi Nishiuchi[10], Hiroshi Ohno[3], Hiroaki Kodama[4], and Naruki Sato[1,2]*

*Affiliations:*
[1]Graduate School of Biology, Faculty of Science, Chiba University, Chiba, Chiba 263-8522, Japan
[2]Graduate School of Science and Engineering, Chiba University, Chiba, Chiba 263-8522, Japan
[3]RIKEN Center for Integrative Medical Sciences, Yokohama, Kanagawa 230-0045, Japan
[4]Graduate School of Horticulture, Chiba University, Matsudo, Chiba 271-8501, Japan
[5]Graduate School of Medical Life Science, Yokohama City University, Yokohama, Kanagawa 230-0045, Japan
[6]Sermas Co., Ltd., Ichikawa, Chiba 272-0033, Japan
[7]Japan Eco-science Co., Ltd., Chiba, Chiba 263-8522, Japan
[8]RIKEN Center for Advanced Photonics, Wako, Saitama 351-0198, Japan
[9]Division of Hematology, Kitasato University School of Allied Health Sciences, Sagamihara, Kanagawa 252-0373, Japan
[10]Division of Integrated Omics Research, Bioscience Core Facility, Research Center for Experimental Modeling of Human Disease, Kanazawa University, Kanazawa, Ishikawa 920-8640, Japan

*Correspondence: Naruki Sato Ph.D.
E-mail: sato@faculty.chiba-u.jp

## Abstract

The gut-muscle axis has been proposed to link gut microbiota with skeletal muscle physiology, yet its universality across livestock species remains unclear. Using aged laying hens — a livestock model with a relatively short digestive tract — we examined the gut microbiota (16S rRNA gene), faecal metabolome, and breast-muscle metabolome by integrative multi-omics analyses in hens fed a *Caldifermentibacillus hisashii*–containing fermented feed or a control diet. Non-metric multidimensional scaling revealed clear separation of the microbial community between groups (stress = 0.0097), characterised by a marked expansion of *Lactobacillus* with the administration of the fermented feed. Variance partitioning showed that the 16S microbiota shared substantial variance with both the faecal (shared $R^2_{adj}$ = 0.54) and muscle (shared $R^2_{adj}$ = 0.48) metabolomes, and partial dbRDA demonstrated that the faecal-to-muscle metabolite association was largely retained after controlling for 16S (direct $R^2 = 0.538 \rightarrow$ partial $R^2 = 0.485$, $p < 0.01$), consistent with faecal metabolites acting as an integral layer linking microbiota to muscle. Cliff's delta–based selection showed depletion of proteolytic taxa and faecal amino acids, and reduced muscle Ornithine and uric acid alongside elevated Hypoxanthine. Because both groups were processed identically post-slaughter, these differences reflect in vivo states: amino acid depletion despite reduced bacterial proteolytic capacity points to enhanced host utilisation, and reduced uric acid — a post-mortem-stable purine end-product in uricotelic chickens — indicates efficient nitrogen turnover rather than accumulation. Collectively, these findings support a putative tripartite model of the gut-muscle axis in aged laying hens, providing a statistically grounded framework for understanding microbial contributions to muscle physiology in aged livestock.

*Abbreviations*: dbRDA, distance-based redundancy analysis

## Introduction

The gut-muscle axis has emerged as a conceptual framework describing bidirectional interactions between the gut microbiota and skeletal muscle physiology [1]. Accumulating evidence from human and rodent studies suggests that gut microbial composition, together with microbially-derived or microbially-modulated metabolites, can influence muscle mass, function, and metabolic status — most prominently exemplified by the link between gut dysbiosis and age-related sarcopenia [2]. Short-chain fatty acids, amino acid derivatives, and other bacterial metabolites are thought to act as molecular mediators of this axis, reaching skeletal muscle through the systemic circulation and modulating energy metabolism, protein turnover, and inflammatory tone [3]. Despite growing interest, however, most studies have focused on humans and rodent models, and whether the gut-muscle axis operates consistently in livestock species — particularly under physiologically distinct conditions such as advanced age — remains poorly characterised. Among livestock, aged laying hens represent a particularly relevant model in this regard: they possess a comparatively short digestive tract, are routinely processed for meat at the end of their laying period, and show considerable individual variation in muscle composition and quality. Understanding how the gut microbiota and its associated metabolites relate to the muscle metabolome in such aged animals could therefore inform both fundamental biology and the practical utilisation of post-laying livestock.

One promising approach to modulate symbiotic microbial communities in livestock is the use of environmental microorganisms, particularly members of the family Bacillaceae [4]. Our group has previously demonstrated that compost-derived thermophilic Bacillaceae can be applied to regulate the microbiota associated with soil and terrestrial plants [5, 6], marine sediments and seagrasses — the latter being a key target of blue-carbon initiatives [7] — as well as insects [8, 9], fish [10-12], livestock [13-16], and rodents [17, 18]. Compost-type fermented feeds containing thermophiles have been shown to contribute to productivity in insects, fish, chickens, and pigs. As an environmentally low-burden approach, the indirect effect of fermented feed on seagrass growth in the inshore zone of an aquaculture farm has also been observed, suggesting that bottom-sediment microbiota are altered following treatment. *Caldifermentibacillus hisashii* (synonyms: *Caldibacillus hisashii*, *Bacillus hisashii*), a major functional thermophilic bacterium present in the fermented feed, is a strong candidate that mimics compost-derived functions. It was originally isolated from an experimental system using germ-free mice implanted with compost [19, 20] as a species closely related to *Bacillus thermoamylovorans* N11 [18], first registered as *B. hisashii* (type strain N-11T = NRBC 110226T = LMG 28201T) [19] and recently reclassified into the genus *Caldifermentibacillus* [21]. Oral administration of *C. hisashii* increases secretory IgA — a key mediator of mucosal immune defence in the gastrointestinal tract [22, 23] — in mice, and in calves it increases the abundance of the phylum Bacteroidetes (syn. Bacteroidota) and reduces the abundance of the methanogenic genus *Methanobrevibacter*, accompanied by improved growth [24]. Collectively, these findings indicate that *Caldifermentibacillus* strains are capable of modulating symbiotic microbial systems across diverse host species, motivating their evaluation in aged livestock where microbial-host interactions may underlie age-related changes in muscle physiology.

Here, we evaluated the relationships among faecal bacterial populations and metabolome, and breast-muscle metabolome in aged laying hens that had been maintained for over two years and either received a *C. hisashii*-containing fermented compost-type feed or a control diet only. Based on integrative multi-omics analyses, such as 16S rRNA gene-based bacterial profiling and GC–MS-based metabolomics of both faeces and muscle, non-metric multidimensional scaling, distance-based redundancy analysis (dbRDA), variance partitioning, partial dbRDA, and Cliff's delta–based feature selection could dissect inter-layer relationships in a statistically grounded manner. The aim was to determine (i) whether feeding the fermented feed differentially shapes gut microbial composition in aged laying hens, (ii) to what extent the gut microbiota co-varies with faecal and muscle metabolite profiles, and (iii) whether these multi-layered relationships can be coherently organised into a putative model of the gut-muscle axis in aged livestock. The results presented in this study describe such a tripartite framework, in which gut microbial differences and their associated faecal metabolites coordinately shape the muscle metabolome, providing a foundation for further mechanistic and applied studies in aged livestock physiology and meat quality.

## Result and Discussion

### *Gut microbial community structure differs between groups*

Non-metric multidimensional scaling (NMDS) based on Bray–Curtis dissimilarity revealed clear separation of the gut microbial community between the Con and Tst groups (stress = 0.0097; Fig. 1a). Samples from the Tst group clustered tightly along the positive side of NMDS1, whereas Con samples were more dispersed in the opposite region, indicating that *C. hisashii* administration substantially restructured the gut microbiota in aged laying hens. The high fidelity of this ordination was confirmed by the Shepard plot ($R^2 = 1$ for both non-metric and linear fits; Fig. 1b).

Compositional analysis at the phylum and genus levels (Fig. 2) showed the basis of this divergence. At the phylum level (Fig. 2a), Bacteroidetes were detectable in all Con samples (approximately 15–35 %) but were almost entirely absent in Tst, while Firmicutes remained the dominant phylum in both groups. At the genus level (Fig. 2b), *Lactobacillus* showed a marked increase in Tst (approximately 60–90 %) compared with Con (approximately 15–60 %), identifying it as a major contributor to the inter-group difference. Although *Lactobacillus* expansion has been widely associated with probiotic-type fermented feed in poultry, the concomitant near-disappearance of Bacteroidetes is a striking feature in this aged-hen model and suggests that *C. hisashii*–containing fermented feed reshapes the gut microbiota at the phylum level more strongly than typical lactic-acid-bacteria interventions.

***Gut microbiota co-varies with faecal and muscle metabolomes***

To examine how the observed microbial restructuring relates to downstream metabolite layers, we performed distance-based redundancy analysis (dbRDA) and variance partitioning (VP) between the 16S microbiota and each metabolome (Fig. 3, Fig. S1). When forward-selected metabolites were used as explanatory variables, the chemistry effect explained substantially more variance in 16S community structure (Fig. S1: $R^2adj = 0.601$ for faecal and 0.771 for muscle metabolites, both $p < 0.01$) than the group effect alone ($R^2adj = 0.463$, $p < 0.01$), indicating that the metabolite layers capture biological variation beyond the simple Con-vs-Tst contrast.

Variance partitioning further demonstrated that the 16S microbiota and metabolome layers share a large fraction of their variation (Fig. 3). For 16S × faecal metabolome (Fig. 3a), the shared component dominated (shared $R^2adj = 0.543$), while the unique contributions of 16S alone (0.012) and faecal metabolites alone (0.127) were small, and the residual was 0.319. The 16S × muscle metabolome comparison (Fig. 3b) showed a similar pattern (shared $R^2adj = 0.480$) but with a larger unique muscle-metabolite component (0.318) and a smaller residual (0.127). Together these results indicate that the gut microbiota and both metabolome layers are tightly co-varying systems, with the muscle metabolome additionally carrying microbiota-independent variation not fully captured by the 16S profile alone.

***Faecal–muscle metabolite linkage***

To address whether the relationship between faecal and muscle metabolites is mediated primarily through the gut microbiota or persists when the microbial profile is statistically controlled for, we examined direct faecal × muscle metabolite associations and applied partial dbRDA (Fig. 4). Variance partitioning between the two metabolome layers in the absence of 16S (Fig. S2) showed a markedly lower shared component (shared $R^2adj = 0.080$) than the 16S-inclusive analyses in Fig. 3, with a sizeable residual (0.573). At first glance, this might suggest that faecal and muscle metabolites are only weakly linked. However, this interpretation overlooks the fact that the variation shared between faecal and muscle metabolites is in large part already absorbed by the 16S microbiota, as shown in Fig. 3.

Partial dbRDA clarified this point (Fig. 4). The direct association between faecal and muscle metabolites — without conditioning — was strong (Direct $R^2 = 0.538$, $p < 0.01$). After statistically partialling out the 16S microbiota, this association was only modestly reduced (Partial $R^2 = 0.485$, $p < 0.01$), a relative decrease of approximately 10 %. In parallel, the variation in muscle metabolites attributable to 16S alone, without considering faecal metabolites, was small and non-significant (Hub-only $R^2 = 0.154$, n.s.). Importantly, this pattern should not be interpreted as the muscle metabolome being independent of the gut microbiota in a biological sense: faecal metabolites are themselves shaped by the microbiota, and the strong faecal–muscle linkage retained after controlling for 16S is most parsimoniously interpreted as the microbial signal having been "carried over" into the faecal metabolite profile, which in turn relates to muscle metabolism. The gut microbiota thus appears to act as an integral, rather than a strictly intermediate, component of this multi-layered relationship in aged laying hens.

***Differentially abundant taxa and metabolites***

To identify the individual features driving the inter-group differences, we applied Cliff's delta–based effect-size analysis to each layer ($|\delta| \geq 0.85$, Mann–Whitney $p < 0.05$; $n = 5$ per group, 10,000 bootstrap iterations). Under the $n = 5$ vs 5 design, the minimum attainable Mann–Whitney p value is 0.0079, corresponding to complete group separation ($|\delta| = 1.0$); most of the features identified here therefore represent near-complete or complete separation between Con and Tst. Because all birds were processed under identical slaughter, sampling, storage, and analytical protocols (see Methods), any post-mortem contribution to metabolite levels is expected to be equivalent between groups; consequently, the inter-group differences described below are interpreted as reflecting in vivo metabolic states rather than processing artefacts.

At the 16S level (Fig. S3), 27 taxa met the criteria. Genera depleted in Tst included *Tissierella*, *Proteiniclasticum*, *Vagococcus*, *Pseudomonas*, *Anaerosphaera*, *Erysipelothrix*, *Ignatzschineria*, and *Gottschalkia*, several of which are anaerobic or facultative anaerobes commonly associated with Bacteroidetes-rich microbiota and with proteolytic or amino acid–fermentative metabolism. Genera enriched in Tst included *Acinetobacter*, *Lachnoclostridium 12*, *Desemzia*, *Kocuria*, *Brevibacterium*, *Atopobium*, *Brachybacterium*, *Actinomyces*, and *Candidatus Stoquefichus*, several of which are aerobic Gram-positive bacteria, including genera commonly associated with fermented feed or environmental sources. This taxon-level pattern is consistent with the macro-scale shift from a Bacteroidetes-containing to a Firmicutes/*Lactobacillus*-dominated community shown in Fig. 2. Notably, *Lactobacillus* itself was not retained by Cliff's delta selection despite its pronounced community-level increase (Fig. 2b). This reflects the conservative nature of the $|\delta| \geq 0.85$ threshold under the $n = 5$ vs 5 design: individual variation in Con (range ≈ 15–60 %) overlapped the lower end of the Tst range (≈ 60–90 %), precluding complete separation. The Cliff's delta–selected taxa therefore represent a complementary set of completely separating, generally lower-abundance features that accompany — rather than recapitulate — the *Lactobacillus*-driven restructuring evident at the compositional level (Fig. 2).

At the faecal metabolome level (Fig. S4), 10 metabolites met the criteria, with five showing complete separation — Ornithine, Alanine, Glycine, Threonine (all amino acids), and Erythrulose-meto-3TMS(2) (a sugar derivative) — each decreasing in Tst. This coordinated depletion of multiple faecal amino acids, considered together with the concomitant reduction of proteolytic and amino acid–fermentative taxa noted above, is informative: a decrease in bacterial proteolytic capacity would not, by itself, be expected to lower faecal amino acid levels through microbial consumption. The most parsimonious interpretation is therefore enhanced host absorption of amino acids across the intestinal epithelium in Tst birds, consistent with the known effects of *Lactobacillus*-enriched microbiota on intestinal barrier integrity and absorptive function.

At the muscle metabolome level (Fig. S5), 14 metabolites met the criteria. Six metabolites decreased in Tst (Cytosine,

Guanine, Uric acid, Ornithine, Adipic acid, Glutaric acid), whereas one metabolite, Hypoxanthine, increased. Two patterns warrant particular attention. First, the elevation of Hypoxanthine together with the depletion of its downstream oxidative products, Guanine and uric acid, is consistent with a reduced flux through xanthine dehydrogenase/oxidase in vivo. Although Hypoxanthine can accumulate post-mortem from ATP degradation, the identical processing of both groups means that this inter-group difference reflects an in vivo state; moreover, uric acid — the terminal product of purine catabolism in uricotelic chickens — is comparatively stable against short-term post-mortem changes, and its reduction therefore provides an independent indicator that nitrogen catabolism is being efficiently processed and excreted rather than accumulated. Second, Ornithine was depleted in both faecal and muscle compartments. Because chickens are uricotelic and the Arginine–Ornithine pathway is a major route of nitrogen handling toward uric acid biosynthesis, the simultaneous reduction of Ornithine in two independent layers — alongside the depletion of other amino acids in faeces — supports active utilisation of nitrogen substrates throughout the gut-muscle axis. The depletion of dicarboxylic acids (Adipic acid, Glutaric acid) similarly indicates efficient progression of fatty acid and amino acid catabolism rather than incomplete oxidation.

***A putative model of the gut-muscle axis in aged laying hens***

Integrating these observations, we propose a putative tripartite model of the gut-muscle axis in aged laying hens (Fig. 4 for the statistical relationships). In this model, administration of *C. hisashii*–containing fermented feed restructures the gut microbiota at the community level (Fig. 1, 2), with the expansion of *Lactobacillus* and near-disappearance of Bacteroidetes accompanied by changes in numerous individual taxa, including a reduction of proteolytic and amino acid–fermentative genera (Fig. S3). Against this background, the depletion of faecal amino acids despite reduced bacterial proteolytic capacity points to enhanced host amino acid absorption (Fig. S4), and the absorbed nitrogen substrates appear to be actively utilised: in muscle, Ornithine and other amino acids are depleted, uric acid is reduced rather than accumulated, and the purine pathway shows a coordinated shift (Hypoxanthine elevation with Guanine and uric acid depletion; Fig. S5). The faecal–muscle metabolite linkage is largely preserved even after the 16S microbiota is statistically controlled for (Fig. 4), indicating that the microbial signal is to a substantial degree embedded within the faecal metabolite layer rather than acting purely as an upstream regulator. Because all birds were processed under identical post-slaughter protocols, these inter-group differences reflect in vivo metabolic states; in particular, uric acid — the terminal product of purine catabolism in uricotelic chickens and comparatively stable against post-mortem changes — provides a robust indicator that nitrogen is being efficiently turned over and excreted rather than accumulated. Collectively, these layers describe a coordinated signature of efficient nitrogen turnover and active metabolism — a pattern more reminiscent of a metabolically active state than of the metabolite accumulation typically associated with aged or sarcopenic muscle, despite the advanced chronological age of the hens studied here. We therefore interpret these convergent observations as raising the possibility of an "anti-aging-like" metabolic response to the fermented feed-mediated gut modulation, while emphasising that this remains a putative interpretation requiring direct validation.

Several limitations should be acknowledged. The sample size (n = 5 vs 5) is small, and although Cliff's delta with bootstrap confidence intervals provides a robust effect-size estimate under such conditions, broader validation in independent cohorts is necessary. The study is observational and correlational, and while partial dbRDA quantifies the statistical interdependence of layers, it does not directly establish causal flow. The metabolite annotations are based on GC–MS profiling, and the biological roles of the highlighted compounds in chicken muscle physiology remain to be experimentally confirmed. Importantly, we did not measure conventional indicators of muscle aging — such as muscle fibre composition, oxidative stress markers, or sarcopenia indices — nor include an age-matched non-aged reference group; the inference of an "anti-aging-like" state is therefore based solely on the convergent multi-omics metabolic profile and requires direct physiological validation. Alternative explanations, including group differences in feed intake or intestinal transit, cannot be fully excluded.

Notwithstanding these limitations, the integrative framework presented here supports the operation of a gut-muscle axis in aged laying hens and provides a statistically grounded foundation for further mechanistic investigation. Given the practical relevance of aged laying hens to meat production and the broader challenge of post-laying livestock utilisation, this putative model may inform future strategies for modulating muscle quality through gut-targeted interventions, and may be relevant to other aged livestock species in which similar microbiota–metabolite–muscle relationships could operate.

## Methods

***Animals and experimental design***

Aged laying hens (over two years old at the end of the experiment) were used in this study. Birds were randomly assigned to two groups (n = 5 per group): a control group (Con; ID140–ID144) maintained on a standard commercial layer diet, and a treatment group (Tst; ID145–ID149) receiving the same diet supplemented with *Caldifermentibacillus hisashii*–containing fermented compost-type feed at 0.05% (w/w) of the basal control diet [13-16]. Housing, lighting regime, water provision, and welfare management followed standard commercial layer husbandry. At the end of the experimental period, faecal samples were collected and the birds were processed; breast muscle (pectoralis major) was excised, immediately frozen in liquid nitrogen, and stored at −80 °C until analysis. All animal procedures were performed in accordance with the relevant institutional and national guidelines.

***16S rRNA gene sequencing and processing***

Genomic DNA was extracted from faecal samples using [DNA extraction kit, manufacturer]. The V4 region of the 16S rRNA gene was amplified and sequenced on the Illumina MiSeq platform. The obtained FASTAQ format files were analysed using QIIME2 (https://qiime2.org) [25]. Taxa were aggregated at the phylum level (D_1 level in SILVA nomenclature) and the genus level (D_5 level in SILVA nomenclature). All rRNA sequence datasets were deposited in the GenBank Sequence Read Archive database as described in the “Data and materials availability” section. For ordination and dbRDA, raw counts were either transformed to relative abundances (Bray–Curtis distance) or subjected to centred log-ratio (CLR) transformation depending on the analysis (see below).

***Bacterial diversity***

The values of α-diversity obtained by Qiime2 were assessed by multiple comparisons as described in the “Statistical analysis” section. The statistical analysis of nonmetric multidimensional scaling (NMDS) values for β-diversity was performed using the “vegan” (distance=bray as the calculation condition) and “pairwiseAdonis” libraries in R software. These were visualized by the following libraries and modules in Python (version 3.10.8) in Mac OS Sequoia (version 15.3) on arm64: “numpy”, “pandas”, “seaborn”, “matplotlib.pyplot”, “statsmodels.formula.api”, “statsmodels.stats.multitest”, “matplotlib.patches”, “matplotlib.lines”.

***Metabolome analysis***

Faecal and breast-muscle metabolites were profiled by gas chromatography–mass spectrometry (GC–MS) [26] . Briefly, the metabolites in the samples were derivatised by methoximation followed by trimethylsilylation (TMS), and was carried out on a Shimadzu GCMS-TQ8030 triple quadrupole mass spectrometer (Shimadzu, Japan) with a capillary column (BPX5) (SGE Analytical Science, Australia). Metabolite abundances were natural-log–transformed prior to multivariate statistical analyses; values $\leq 0$ were treated as missing.

***Macroscopic community analysis (NMDS)***

Macroscopic differences in microbial community structure between groups were assessed by non-metric multidimensional scaling (NMDS) based on Bray–Curtis dissimilarity, computed from genus-level relative abundances. NMDS ordination was performed using the metaMDS function in the vegan R package with Bray–Curtis dissimilarity, two dimensions, and multiple random starts to ensure convergence. Group-wise 95 % confidence ellipses were overlaid on the NMDS ordination (Fig. 1a) using ggplot2::stat_ellipse. Ordination quality was assessed by the stress value, and the fidelity between observed dissimilarities and ordination distances was evaluated using a Shepard plot (Fig. 1b).

***Distance-based redundancy analysis (dbRDA) and variance partitioning***

Constrained ordination was performed by distance-based redundancy analysis (dbRDA) using the capscale function in the vegan R package, with Bray–Curtis dissimilarity for count-based 16S data (relative abundances) and Euclidean distance for log-transformed metabolome data. Two types of explanatory variables were considered: (i) the group factor (cnd; Con vs Tst), and (ii) selected continuous variables from the metabolome layers ("chemistry").

Owing to the small sample size relative to the number of candidate variables ($p \gg n$ in the original metabolome matrices), variable selection was performed by forward selection using the ordistep function (10,000 permutations, $p < 0.05$ for retention) rather than full-model variance inflation factor (VIF) screening, to avoid overfitting. Statistical significance of each constrained ordination was tested by permutation-based ANOVA (anova.cca, 999 permutations), and adjusted $R^2$ ($R^2$adj) was obtained using RsquareAdj.

Variance partitioning between the 16S microbiota and each metabolome layer was performed using the varpart function in vegan. For each pair, four fractions were estimated: the unique contribution of factor 1, the unique contribution of factor 2, their shared component, and the residual. All values are reported as adjusted $R^2$. Layer-wise comparisons across (16S × faecal metabolome), (16S × muscle metabolome), and (faecal × muscle metabolome) were performed in parallel to allow direct comparison.

***Partial dbRDA for inter-layer relationship***

To examine whether the relationship between two layers persists after statistically controlling for a third layer (referred to here as the "hub"), partial dbRDA was performed. For the faecal–muscle metabolite relationship, four models were compared: 1) Direct model: muscle metabolome ~ faecal metabolites (Bray–Curtis or Euclidean distance, see above). 2) Partial model with 16S as hub: muscle metabolome ~ faecal metabolites + Condition(16S), where 16S enters as a conditioning matrix. 3) Hub-only model: muscle metabolome ~ 16S. 4) Reciprocal model (used for sanity checking): muscle metabolome ~ 16S + Condition (faecal metabolites). Each model was fitted using capscale and tested by permutation ANOVA (999 permutations). Adjusted $R^2$ was extracted as $R^2$adj. The relative change between Direct $R^2$ and Partial $R^2$ (after controlling for the hub) was used to assess the proportion of variance explained that was attributable to the hub.

## *Cliff's delta–based feature selection*

To identify individual features (taxa or metabolites) differing between Con and Tst, Cliff's delta effect size was computed for each feature, complemented by the Mann–Whitney U test (two-sided) for statistical significance and Welch's t test for parametric reference.

For each feature, Cliff's delta analysis was calculated to compare the two groups. A 95 % confidence interval was obtained by bootstrap resampling (10,000 iterations). Effect size magnitudes were classified following Romano *et al.* (2006): negligible ($|\delta| < 0.147$), small ($|\delta| < 0.330$), medium ($|\delta| < 0.474$), and large ($|\delta| \geq 0.474$).

Features were considered differentially abundant when (i) Mann–Whitney $p < 0.05$ and (ii) |Cliff's $\delta| \geq 0.85$, the latter corresponding to near-complete or complete group separation under the $n = 5$ vs 5 design (minimum attainable Mann–Whitney $p = 0.0079$). For the faecal and muscle metabolomes, log-transformed values were used; for 16S data, both relative abundances and CLR-transformed values were considered, with the latter used for parametric analyses. Cliff's delta is rank-based and therefore invariant to monotonic transformations such as log or CLR.

Selected features were visualised as bubble charts with Cliff's delta on the x-axis, feature name on the y-axis, bubble size proportional to $-\log_{10}(p)$, bubble colour indicating direction of change (red = increase in Tst, blue = decrease), and bootstrap confidence intervals shown as horizontal lines.

***Statistical environment and reproducibility***

All multivariate statistical analyses (NMDS, dbRDA, variance partitioning, partial dbRDA) were performed using the vegan package of R software. Cliff's delta analyses and visualisations were performed in Python using numpy, scipy, pandas, and matplotlib. Sample identifiers, raw input matrices, analysis scripts, and reproducible parameter settings (including random seeds for bootstrap resampling) are provided in the Supplementary Materials. An AI-assisted tool (Claude, Anthropic) was used to support code development for statistical analysis and visualisation.

## Acknowledgments

We are grateful to Mr. Toshiyuki Ohyama (Miyakoji Farm Co., Ltd.) for help with the preparation of the samples.

## Author contributions

Naruki Sato, Hirokuni Miyamoto, Hiroaki Kodama conceptualised the experiments; Hirokuni Miyamoto performed the adjustments of the sampling; Arisa Sano, Daisuke Yamamoto, Tsubasa Irie, Karin Suzuki, Yu Adachi, Aoi Fukushima, Shigeru Moriya, Makiko Matsuura, Naoko Tsuji, Takashi Satoh, Tamotsu Kato performed the experiments to analyze; Arisa Sano, Daisuke Yamamoto, Tsubasa Irie, Karin Suzuki, Yu Adachi, Aoi Fukushima, Hirokuni Miyamoto analysed the raw data; Hirokuni Miyamoto, Shigeharu Moriya, Hiroshi Ohno, Hiroaki Kodama, Narukuki Sato provided analytical methods and/or equipments; Hirokuni Miyamoto, Hiroshi Ohno, Hiroaki Kodama, Naruki Sato supervised the study; Yu Adachi, Hirokuni Miyamoto, and Naruki Sato wrote and reviewed the manuscript. All the authors have read and approved the final manuscript.

## Data availability

Raw files of the bacterial 16S rRNA (V4) data were deposited in the DNA Data Bank of Japan (DDBJ) under NCBI BioProject accession number PRJDBXXXXX (BioSample Accession no. SAMDXXXX-SAMDXXXX). All the data were stored in a source data file. The source commands will be provided on the GitHub sites. Please contact the corresponding authors to obtain any additional information. If you need the information not listed here, please contact the corresponding authors.

## Funding

This work was partly supported by a grant from the Subsidy for Livestock Promotion, Japan Racing Association (JRA).

## Competing interests

The authors declare no competing interests.

## Reference

1. Lustgarten MS. The Role of the Gut Microbiome on Skeletal Muscle Mass and Physical Function: 2019 Update. Front Physiol. 2019;10:1435
2. Picca A, Fanelli F, Calvani R, Mule G, Pesce V, Sisto A, et al. Gut Dysbiosis and Muscle Aging: Searching for Novel Targets against Sarcopenia. Mediators Inflamm. 2018;2018:7026198
3. Frampton J, Murphy KG, Frost G, Chambers ES. Short-chain fatty acids as potential regulators of skeletal muscle metabolism and function. Nat Metab. 2020;2:840-8
4. Niisawa C, Oka S, Kodama H, Hirai M, Kumagai Y, Mori K, et al. Microbial analysis of a composted product of marine animal resources and isolation of bacteria antagonistic to a plant pathogen from the compost. J Gen Appl Microbiol. 2008;54:149–58
5. Ishikawa K, Ohmori T, Miyamoto H, Ito T, Kumagai Y, Sonoda M, et al. Denitrification in soil amended with thermophile-fermented compost suppresses nitrate accumulation in plants. Appl Microbiol Biotechnol. 2013;97:1349–59
6. Miyamoto H, Shigeta K, Suda W, Ichihashi Y, Nihei N, Matsuura M, et al. An agroecological structure model of compost-soil-plant interactions for sustainable organic farming. ISME Commun. 2023;3:28
7. Miyamoto H, Kawachi N, Kurotani A, Moriya S, Suda W, Suzuki K, et al. Computational estimation of sediment symbiotic bacterial structures of seagrasses overgrowing downstream of onshore aquaculture. Environ Res. 2023;219:115130
8. Miyamoto H, Asano F, Ishizawa K, Suda W, Miyamoto H, Tsuji N, et al. A potential network structure of symbiotic bacteria involved in carbon and nitrogen metabolism of wood-utilizing insect larvae. Sci Total Environ. 2022;836:155520
9. Asano F, Tsuboi A, Moriya S, Kato T, Tsuji N, Nakaguma T, et al. Amendment of a thermophile-fermented compost to humus improves the growth of female larvae of the Hercules beetle *Dynastes hercules* (Coleoptera: scarabaeidae). J Appl Microbiol. 2023;134:lxac006
10. Tanaka R, Miyamoto H, Kodama H, Kawachi N, Udagawa M, Miyamoto H, et al. Feed additives with thermophile-fermented compost enhance concentrations of free amino acids in the muscle of the flatfish *Paralichthys olivaceus*. J Gen Appl Microbiol. 2010;56:61–5
11. Tanaka R, Miyamoto H, Inoue S, Shigeta K, Kondo M, Ito T, et al. Thermophile-fermented compost as a fish feed additive modulates lipid peroxidation and free amino acid contents in the muscle of the carp, *Cyprinus carpio*. J Biosci Bioeng. 2016;121:530–5
12. Miyamoto H, Ito S, Suzuki K, Tamachi S, Yamada S, Nagatsuka T, et al. A putative research model for sustainable fisheries driven by noninvasive diagnostic imaging. The Innovation Life. 2025;3
13. Ito T, Miyamoto H, Kumagai Y, Udagawa M, Shinmyo T, Mori K, et al. Thermophile-fermented compost extract as a possible feed additive to enhance fecundity in the laying hen and pig: modulation of gut metabolism. J Biosci Bioeng. 2016;121:659–64
14. Miyamoto H, Kodama H, Udagawa M, Mori K, Matsumoto J, Oosaki H, et al. The oral administration of thermophile-fermented compost extract and its influence on stillbirths and growth rate of pre-weaning piglets. Res Vet Sci. 2012;93:137–42
15. Inabu Y, Miyamoto H, Takahashi H, Kato T, Moriya S, Kurotani A, et al. Compost fermented with thermophilic Bacillaceae reduces heat stress-induced mortality in laying hens through gut microbial modulation. Anim Microbiome. 2026;8:9
16. Yoshikawa S, Itaya K, Hoshina R, Tashiro Y, Suda W, Cho Y, et al. Thermophile-fermented feed modulates the gut microbiota related to lactate metabolism in pigs. Journal of Applied Microbiology. 2024; doi:10.1093/jambio/lxae254.
17. Satoh T, Nishiuchi T, Naito T, Matsushita T, Kodama H, Miyamoto H, et al. Impact of oral administration of compost extract on gene expression in the rat gastrointestinal tract. J Biosci Bioeng. 2012;114:500–5
18. Miyamoto H, Shimada E, Satoh T, Tanaka R, Oshima K, Suda W, et al. Thermophile-fermented compost as a possible scavenging feed additive to prevent peroxidation. J Biosci Bioeng. 2013;116:203–8
19. Miyamoto H, Seta M, Horiuchi S, Iwasawa Y, Naito T, Nishida A, et al. Potential probiotic thermophiles isolated from mice after compost ingestion. J Appl Microbiol. 2013;114:1147–57
20. Nishida A, Miyamoto H, Horiuchi S, Watanabe R, Morita H, Fukuda S, et al. *Bacillus hisashii* sp. nov., isolated from the caeca of gnotobiotic mice fed with thermophile-fermented compost. Int J Syst Evol Microbiol. 2015;65:3944–9
21. Gupta RS, Patel S, Saini N, Chen S. Robust demarcation of 17 distinct *Bacillus* species clades, proposed as novel *Bacillaceae* genera, by phylogenomics and comparative genomic analyses: description of *Robertmurraya kyonggiensis* sp. nov. and proposal for an emended genus *Bacillus* limiting it only to the members of the Subtilis and Cereus clades of species. Int J Syst Evol Microbiol. 2020;70:5753–98
22. Mantis NJ, Rol N, Corthesy B. Secretory IgA's complex roles in immunity and mucosal homeostasis in the gut. Mucosal Immunol. 2011;4:603-11
23. Takeuchi T, Miyauchi E, Kanaya T, Kato T, Nakanishi Y, Watanabe T, et al. Acetate differentially regulates IgA reactivity to commensal bacteria. Nature. 2021;595:560-4
24. Inabu Y, Taguchi Y, Miyamoto H, Etoh T, Shiotsuka Y, Fujino R, et al. Development of a novel feeding method for Japanese black calves with thermophile probiotics at postweaning. J Appl Microbiol. 2022;132:3870–82
25. Bolyen E, Rideout JR, Dillon MR, Bokulich NA, Abnet CC, Al-Ghalith GA, et al. Reproducible, interactive, scalable and extensible microbiome data science using QIIME 2. Nat Biotechnol. 2019;37:852–7
26. Nishiumi S, Shinohara M, Ikeda A, Yoshie T, Hatano N, Kakuyama S, et al. Serum metabolomics as a novel diagnostic approach for pancreatic cancer. Metabolomics. 2010;6:518-28

**Figure legends**

**Fig.1**

**Macroscopic differences in gut microbiota community structure between Control (Con) and Treatment (Tst) groups**

(a) Non-metric multidimensional scaling (NMDS) ordination based on 16S rRNA gene composition (n = 5 per group). Each point represents one sample; shape indicates group (squares = Con, circles = Tst) and colour indicates individual animal ID (ID140–ID149). Group-wise 95% confidence ellipses are shown. Distances were computed using the Bray–Curtis dissimilarity index on relative abundance data; stress value is indicated on the plot. (b) Shepard plot showing observed Bray–Curtis dissimilarity (x-axis) versus ordination distance in NMDS space (y-axis). Non-metric and linear fit $R^2$ values are indicated on the plot.

**Fig.2**

**Compositional differences in gut microbiota between Con and Tst groups at the phylum and genus levels**

Stacked bar charts of 16S rRNA gene-based relative abundance for each individual sample (Con: ID140–ID144; Tst: ID145–ID149). Only taxa with mean relative abundance ≥ 10% are coloured; remaining taxa are pooled as "Others" (grey). (a) Phylum-level composition (Firmicutes, dark orange; Bacteroidetes, light orange; Others, grey). (b) Genus-level composition highlighting *Lactobacillus* (orange) versus Others (grey).

**Fig.3**

**Variance partitioning between the 16S microbiota and each metabolome layer**

Variance partitioning based on distance-based redundancy analysis (dbRDA), showing the proportion of variation in the 16S community (Bray–Curtis dissimilarity) explained by the 16S set, the metabolome set, their shared component, and residual variation. Values are adjusted $R^2$ ($R^2$adj) from the vegan::varpart framework. Forward-selected metabolites (see Fig. S1) were used as the chemistry term. (a) 16S × faecal metabolites. (b) 16S × muscle metabolites.

**Fig.4**

**Partial dbRDA framework evaluating the faecal–muscle metabolite relationship with respect to the 16S microbiota**

Boxes denote the three data layers: 16S, the 16S rRNA gene–based gut microbiota; FecMetab, the faecal metabolome; and MusMetab, the breast-muscle metabolome. All statistics shown derive from the same partial dbRDA analysis and are expressed as adjusted $R^2$ ($R^2$adj). The solid green arrow indicates the faecal → muscle metabolite association: Direct $R^2$, the unconditioned association between FecMetab and MusMetab; Partial $R^2$, the same association after statistically conditioning on the 16S microbiota. The solid grey arrow indicates the Hub-only $R^2$, the variance in MusMetab explained by 16S alone (16S → muscle, without faecal metabolites). The dashed grey arrow (16S → FecMetab) denotes the biologically established relationship that faecal metabolites are shaped by the gut microbiota; this relationship is quantified separately by variance partitioning (Fig. 3a) and is shown here without a value to keep all on-diagram statistics within a single analytical framework. Significance codes: ** $p < 0.01$; n.s., not significant.

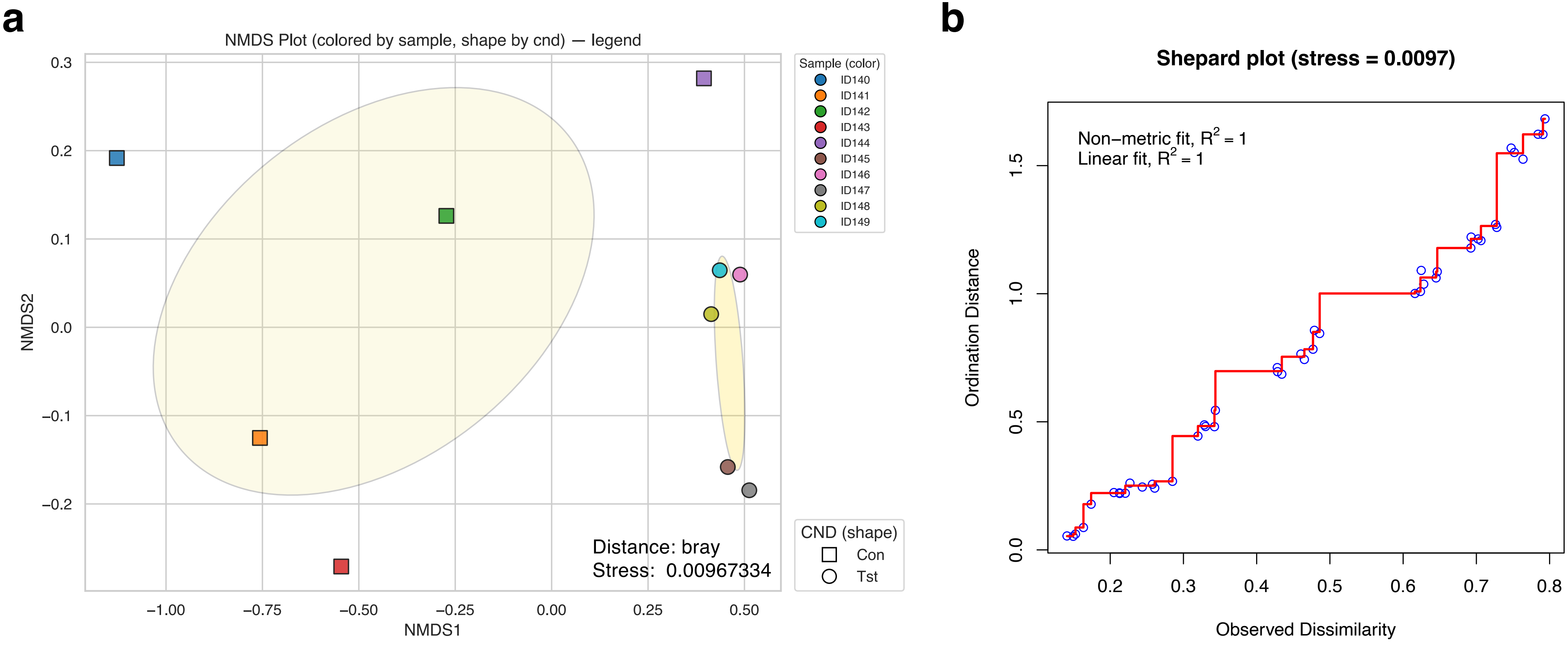
a
NMDS Plot (colored by sample, shape by cnd) — legend
NMDS2
NMDS1
Distance: bray
Stress: 0.00967334
Sample (color)
ID140
ID141
ID142
ID143
ID144
ID145
ID146
ID147
ID148
ID149
CND (shape)
Con
Tst
b
Shepard plot (stress = 0.0097)
Non−metric fit, R2 = 1
Linear fit, R2 = 1
Ordination Distance
Observed Dissimilarity

Fig.1

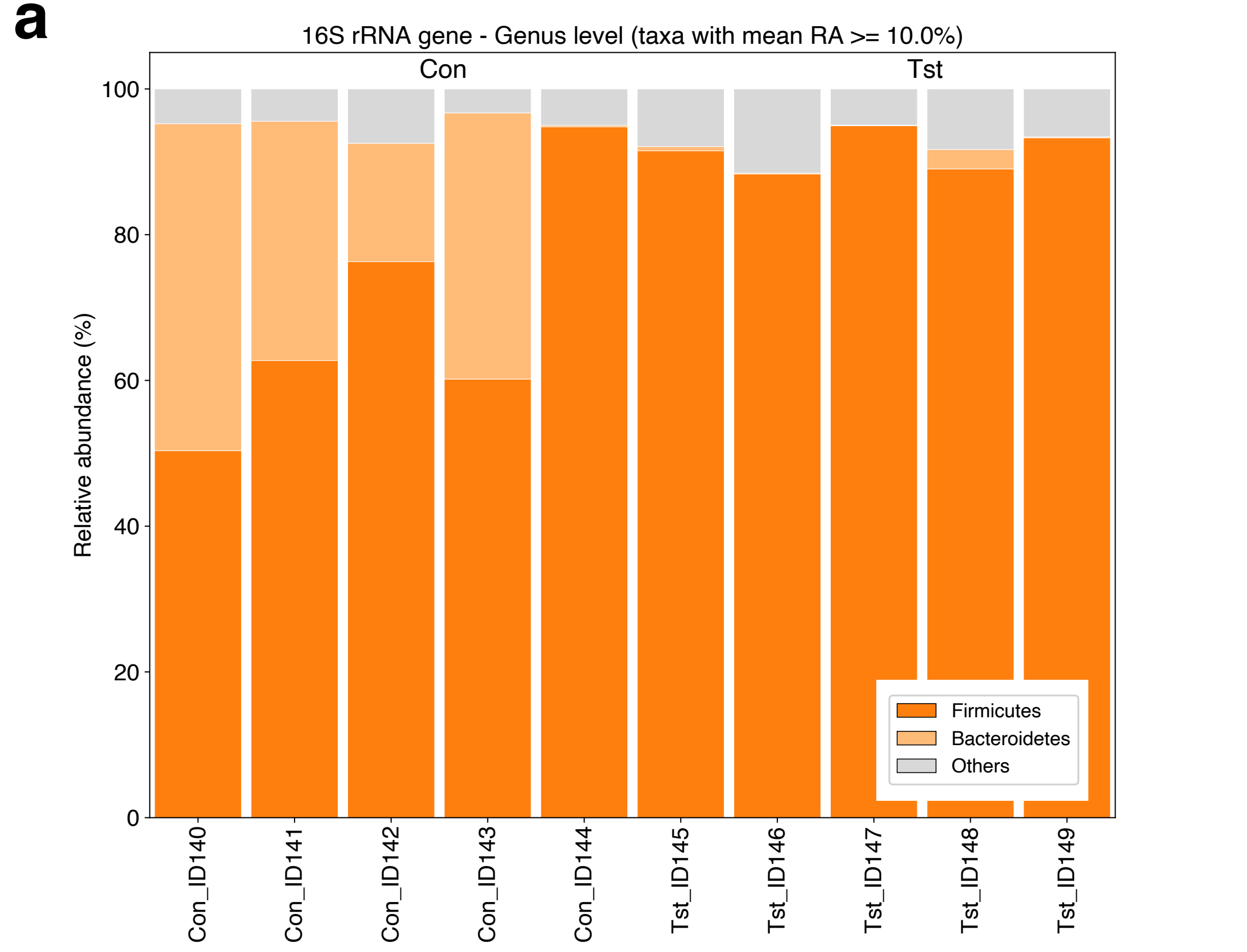
a
16S rRNA gene - Genus level (taxa with mean RA >= 10.0%)
Con
Tst
Relative abundance (%)
100
80
60
40
20
0
Firmicutes
Bacteroidetes
Others
Con_ID140
Con_ID141
Con_ID142
Con_ID143
Con_ID144
Tst_ID145
Tst_ID146
Tst_ID147
Tst_ID148
Tst_ID149

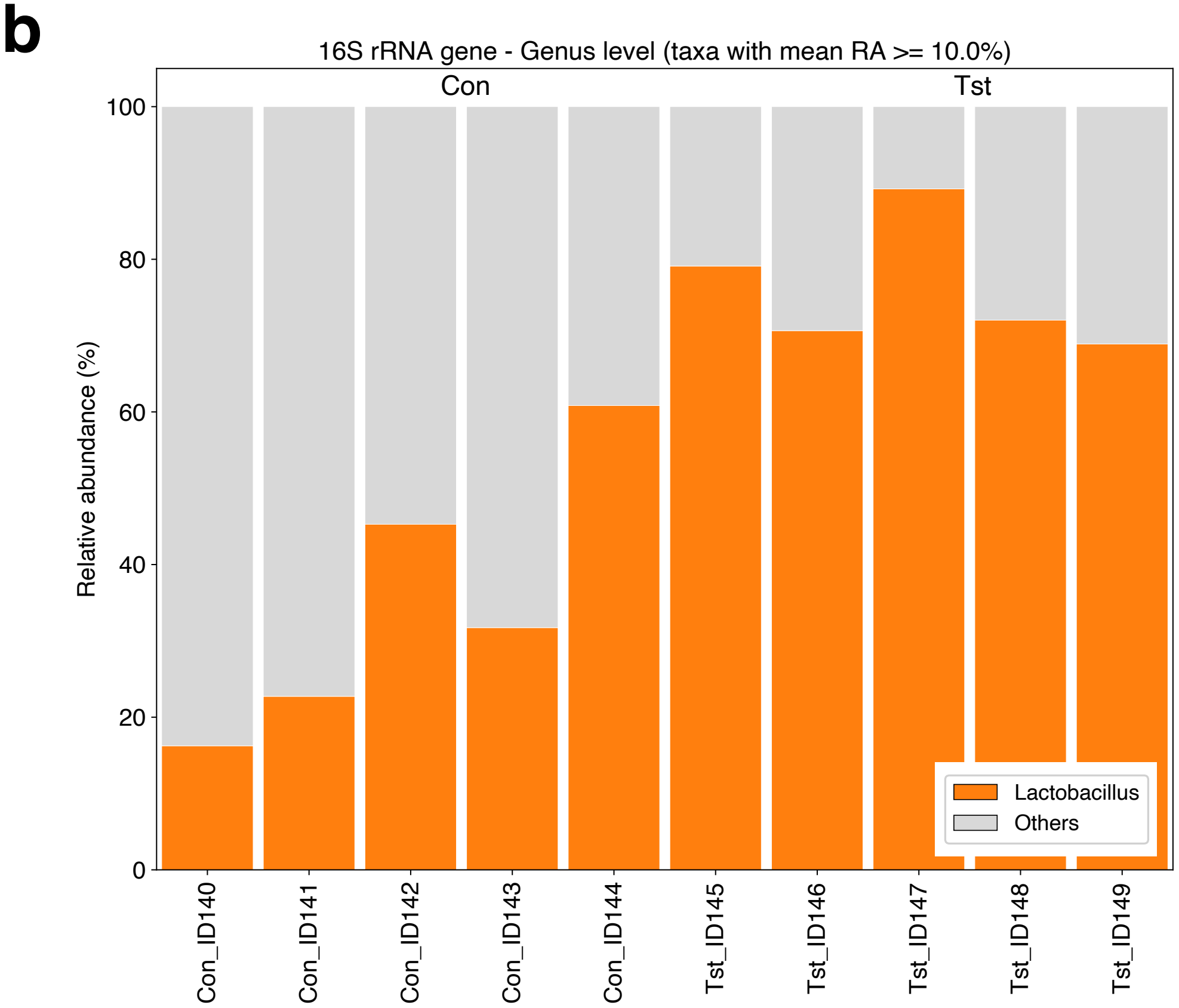
b
16S rRNA gene - Genus level (taxa with mean RA >= 10.0%)
Con
Tst
Relative abundance (%)
100
80
60
40
20
0
Lactobacillus
Others
Con_ID140
Con_ID141
Con_ID142
Con_ID143
Con_ID144
Tst_ID145
Tst_ID146
Tst_ID147
Tst_ID148
Tst_ID149

**Fig.2**

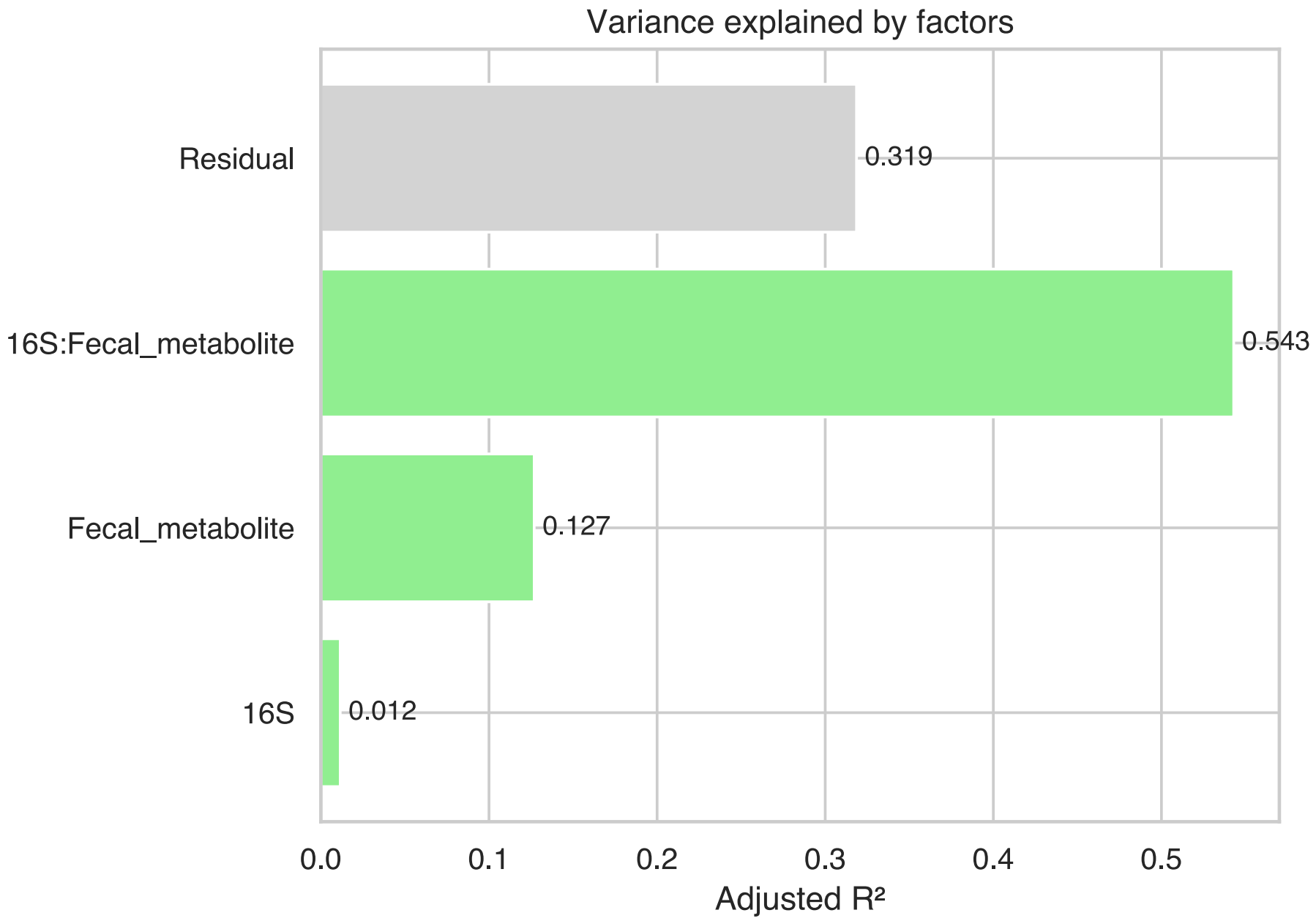

a
Variance explained by factors
Residual
16S:Fecal_metabolite
Fecal_metabolite
16S
0.319
0.543
0.127
0.012
0.0
0.1
0.2
0.3
0.4
0.5
Adjusted R²

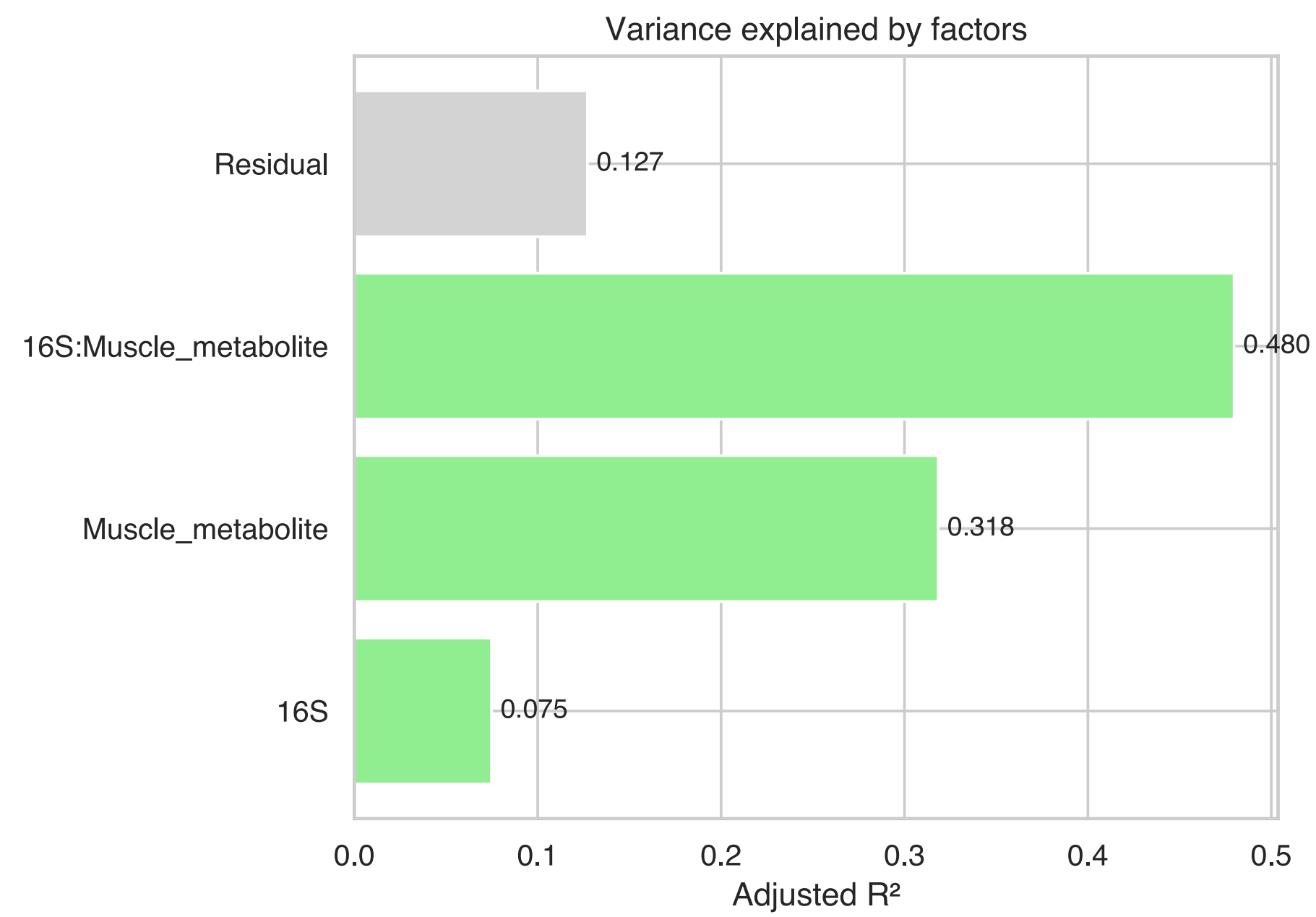

b
Variance explained by factors
Residual
16S:Muscle_metabolite
Muscle_metabolite
16S
0.127
0.480
0.318
0.075
0.0
0.1
0.2
0.3
0.4
0.5
Adjusted R²

**Fig.3**

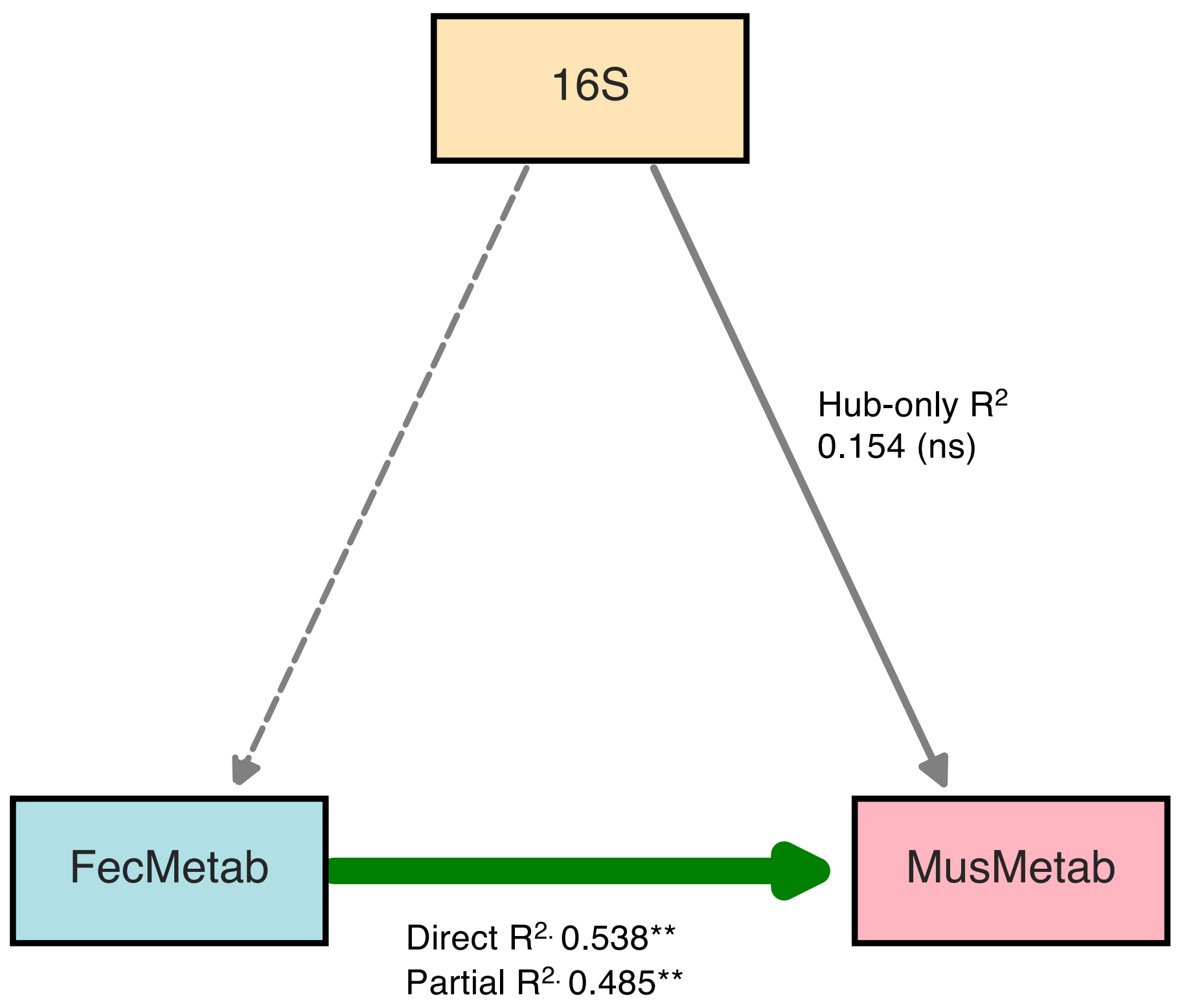
16S
Hub-only R2
0.154 (ns)
FecMetab
MusMetab
Direct R2. 0.538**
Partial R2. 0.485**

**Fig.4**

# Supplementary Information

# A putative model of the gut-muscle axis in aged livestock

Correspondence: Naruki Sato Ph.D.
E-mail: sato@faculty.chiba-u.jp

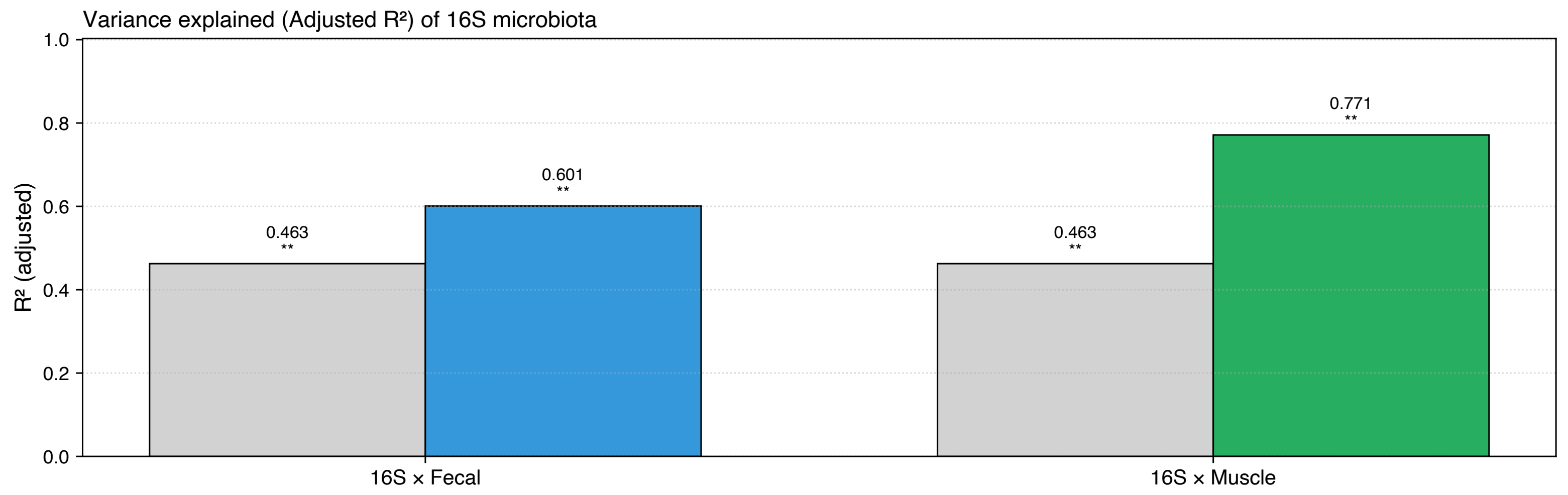

**Figure S1.**

**Adjusted $R^2$ of the 16S microbiota explained by the group effect and forward-selected metabolites**

Bar plots comparing two predictors of 16S community structure (Bray–Curtis dissimilarity) for each metabolome layer (16S × faecal, 16S × muscle): the group effect (cnd; Con vs Tst; grey bars) and forward-selected metabolites (coloured bars). Values are adjusted $R^2$ ($R^2$adj) from dbRDA with permutation-based significance. The group effect is identical across layers because it is computed from the 16S community and group factor alone, independent of the metabolite layer. Significance codes: ** $p < 0.01$.

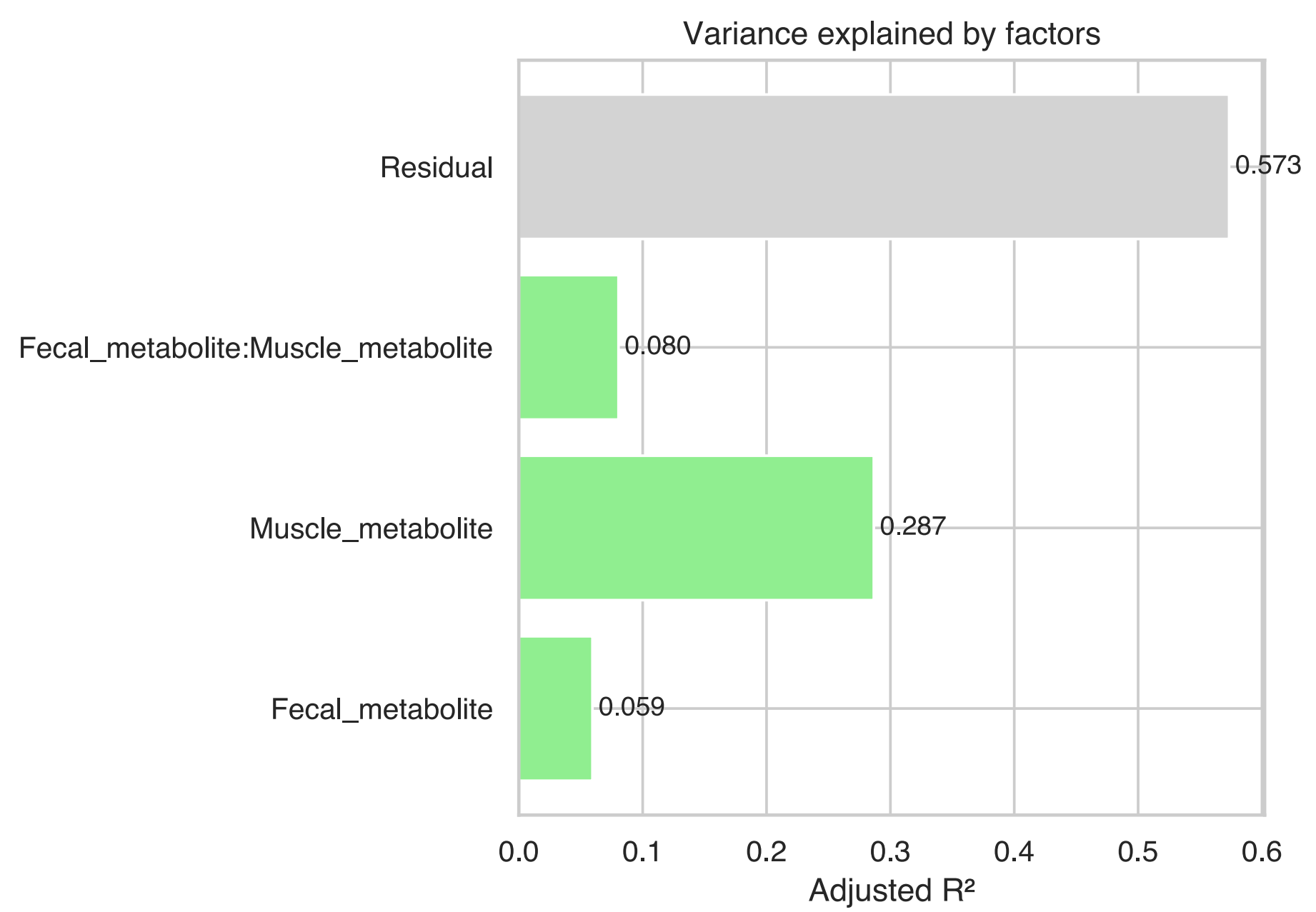

**Figure S2**

**Fecal–muscle metabolite relationship**

Variance partitioning between faecal and muscle metabolites (16S not included in the model), showing unique, shared, and residual components as adjusted $R^2$ ($R^2$adj).

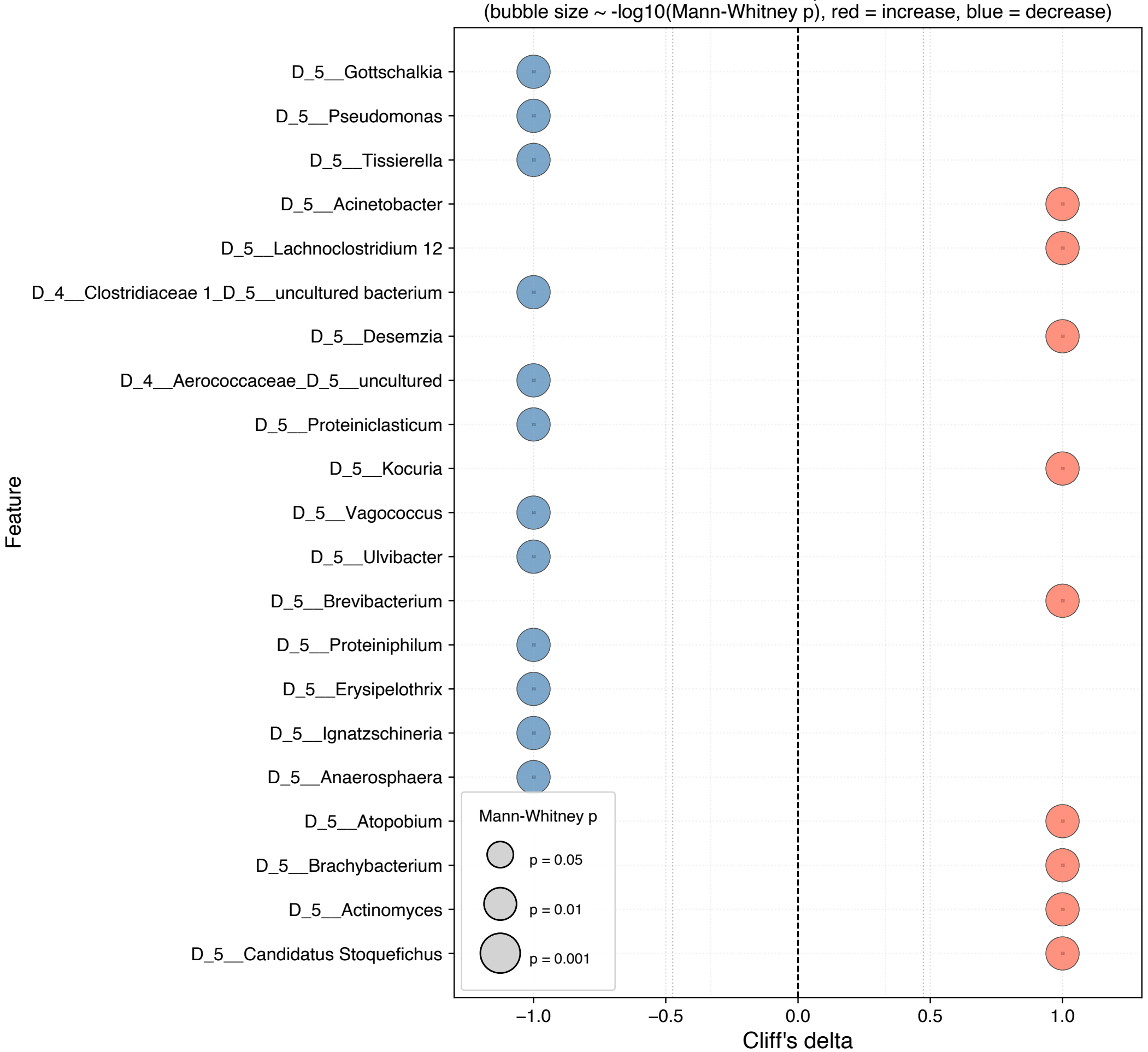

**Figure S3**

**Cliff's delta effect sizes for genus-level 16S taxa**

Bubble chart of Cliff's delta (δ) for taxa meeting the selection criteria (|δ| ≥ 0.85, Mann–Whitney $p < 0.05$; n = 5 per group; 10,000 bootstrap iterations for 95% CI). The x-axis shows Cliff's δ (positive = higher in Tst; negative = higher in Con). Bubble size is proportional to $-\log_{10}$(Mann–Whitney p); colour denotes direction (red = increase in Tst, blue = decrease). Horizontal lines indicate 95% bootstrap confidence intervals.

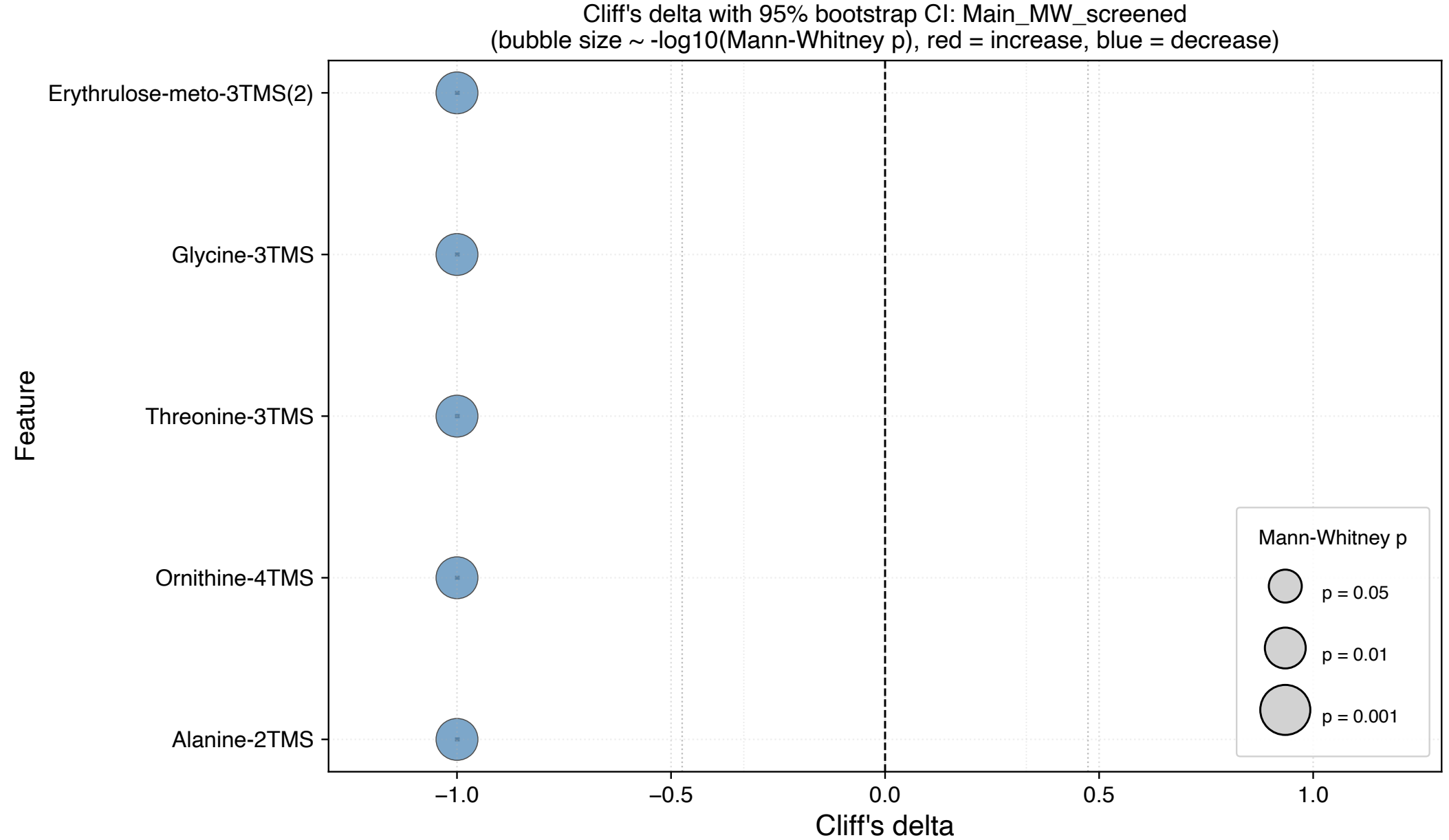

**Figure S4**

**Cliff's delta effect sizes for faecal metabolites**

Bubble chart constructed as in Fig. S3 ($|\delta| \geq 0.85$, Mann–Whitney $p < 0.05$; $n = 5$ per group). Axes, bubble size, colour, and confidence intervals are defined as in Fig. S2.

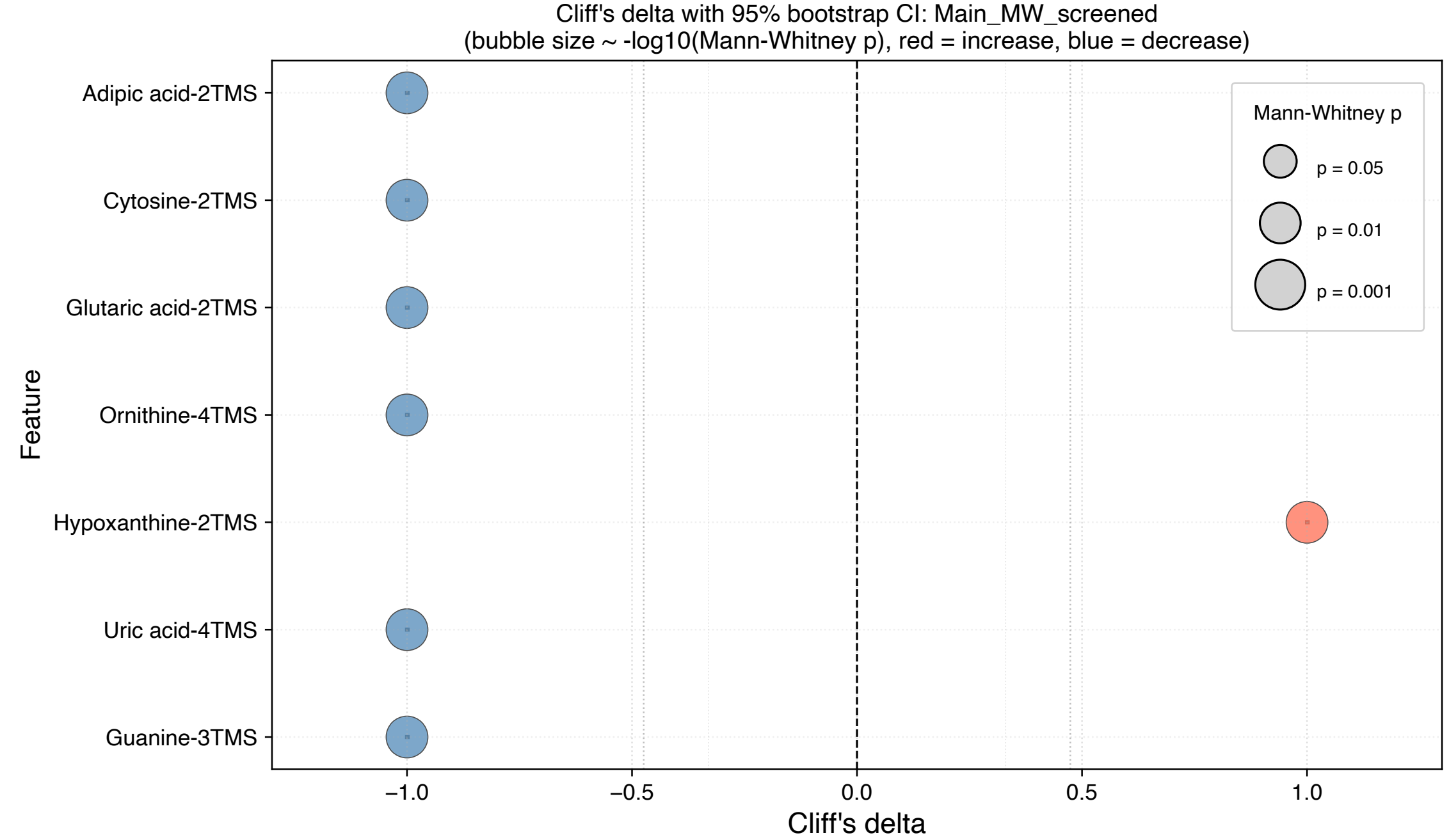

**Figure S5**

**Cliff's delta effect sizes for muscle metabolites**

Bubble chart constructed as in Fig. S3 (|δ| ≥ 0.85, Mann–Whitney $p < 0.05$; $n = 5$ per group). Axes, bubble size, colour, and confidence intervals are defined as in Fig. S2.